# Near orthogonal launch of SPR modes in Au films

John Canning,[1] Akib Karim,[1] Nikki Tzoumis,[1] Yanzhen Tan,[1,2] Rodolfo Patyk,[1,3] Brant C. Gibson[4]

[1] interdisciplinary Photonics Laboratories, School of Chemistry, The University of Sydney, NSW 2006, Australia
[2] Institute of Photonics Technology, Jinan University, Guangzhou, 510632, China
[3] Graduate School of Electrical Engineering & Applied Computer Science, Federal University of Technology - Paraná, 80230-901 Curitiba PR, Brazil
[4] Chemical & Quantum Physics, School of Applied Sciences, RMIT University, VIC 3001, Australia
*Corresponding author: john.canning@sydney.edu.au

We report the excitation of a surface plasmon resonance (SPR) close to the orthogonal axis of a gold (Au) film on borosilicate glass. Direct spectroscopic measurement of SPR shifts using different liquids are made at ~5° incidence within a reflection spectrophotometer. Scattering of light that is able to penetrate across the film at the interfaces is the proposed mechanism by which coupling, and plasmon localization, is established. © 2013 Optical Society of America

Whilst surface wave phenomena at metal-dielectric interfaces arguably stretch back to Wood [1] who noted anomalous reductions in scattered light within metal gratings many decades ago, and coupling into these have been studied for some time [2-5], it has relatively recently received broad attention through extensive applications to sensing, particularly biosensing [6]. In most film cases excitation of the surface wave has proven challenging and generally has required phase matching through an evanescent near field and the electron cloud at the interface. This has been done most commonly using prisms [3,4], end coupling using various optical wave guiding configurations [7,8] and diffractive grating configurations [9-11] which hark back to Wood's observations.

The typical excitation condition of an SPR within a thin metal film is described as being dependent on phase matching the propagation of a surface wave within a medium with both real and imaginary parts of the complex permittivity, or dielectric constant, with that of an in-plane photon excitation source usually in a non-magnetic dielectric such as a silicate glass. The real and imaginary components of the dielectric constant associated with a metal in particular relate the physically distinct real parts of the refractive index (negative with respect to the dielectric) and the imaginary parts which, in metals, can involve energy loss through displacement (current) of the electron plasma. The complex dielectric is described as a sum of the two parts:

$$\tilde{\epsilon} = \epsilon' + i\epsilon'' = (n + i\kappa)^2 \quad (1)$$

Light of *p*-polarisation impinging on a metal from the silica side will generate a time dependent polarization charge at the interface whereas *s*-polarised light does not. The plasma does not have an instantaneous response in practice, since the electrons have a finite mass and scatter energy as phonons or through defect states producing the resistive scattering measured as Ohmic loss (or Joule heating). The two-dimensional nature of the surface also constricts perfect oscillations. Therefore, many surface phenomena are associated with appropriate coupling conditions for optimal energies below the plasma frequency where electrons can respond and dispersion is sufficiently high.

The wave vector of a surface plasmon at the metal-dielectric interface is related to the incident photon vector, $k_0$, by:

$$k_{SPR} = k_0 \left(\frac{\epsilon_1' \epsilon_2'}{\epsilon_1' + \epsilon_2'}\right)^{\frac{1}{2}} \quad (2)$$

where $\epsilon_1'$ is the real part of the dielectric constant of the metal and $\epsilon_2' = n_2^2$ is the constant of the dielectric. From this expression, it would appear the real contribution of equation (1) determines coupling and it's clear the propagation constant of the SPR is always larger than the photon constant. For the typical situation involving a Au film on borosilicate glass, the calculated SPR coupling angle will be $\theta \sim 44°$ to the orthogonal axis.

As for many materials with negligible imaginary components, if the real part of the dielectric constant is considered, there is a familiar Brewster condition established where *p*-polarised light can in principle transmit through a thin film whilst *s*-polarised light is reflected: $\theta = tan^{-1}(n_3/n_1)$ where $n_3$ is the refractive index of the initial dielectric medium light is propagating in. In bulk metals, absorption and Ohmic loss complicates the condition and no typical Brewster transmission is observed. When the film thickness is on the order used to generate plasmons, the exact description appears more complicated than is usually considered because the photon length scale is significantly larger than the film thickness. There will be transmission of some *p*-polarized light and this was observed both at the SPR condition (as a mix of *p* and *s* seemingly indicative of absorption and re-emission) and through the film at the Brewster condition where the real part of equation (1) is met. Considering Au

(at 635 nm is $n_1 = 0.178$ [12]) on borosilicate glass ($n_3 \sim 1.45$), the Brewster angle is $\theta \sim 7°$ at 635 nm. Although launching at this angle might be considered the optimal way to maximize light at the other side of the film compared to an exponentially decaying evanescent field, at this particular condition phase matching is not met and there is no SPR generation in a uniform film.

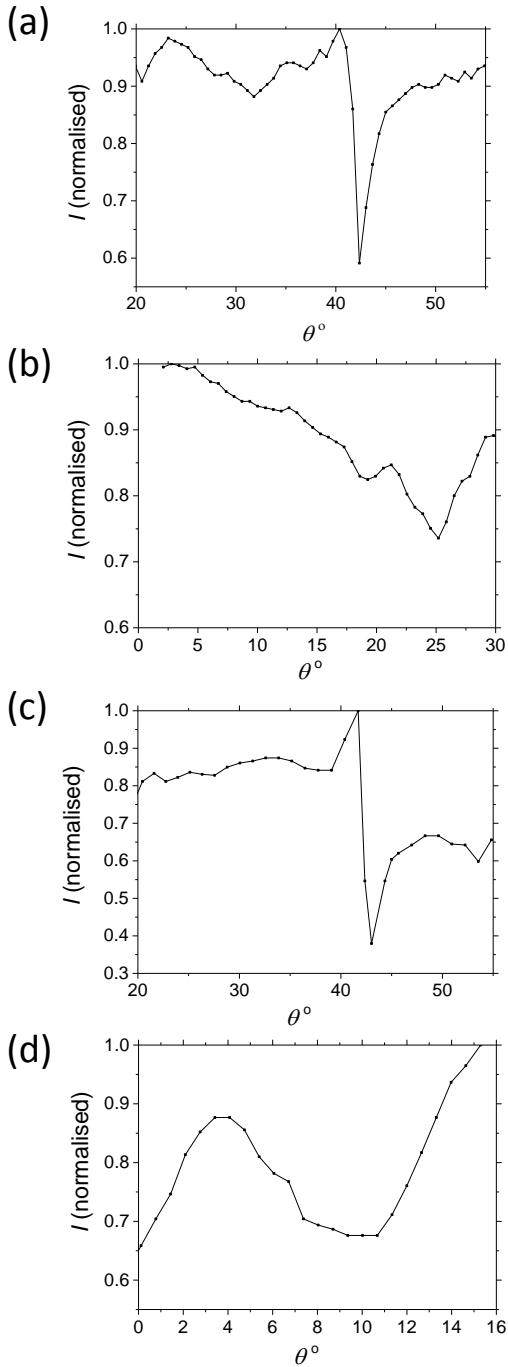

Fig. 1. Reflected signal at 632.5 nm vs angle of incidence, $\theta$, from the orthogonal axis to the base of the borosilicate slide within the Kretchmann setup: (a) & (c) the classical sharp SPR for the 45 nm in-house film and 50 nm PHASIS commercial film respectively; (b) & (d) the secondary, broader feature at smaller angles away from the orthogonal axis.

We speculated whether a non-uniform film would lead to omnidirectional scattering of the $p$-transmitted light at or near the Brewster condition such that some quasi-phase matching condition might be generated by the time the light reaches the far-side. In this way reflected light is avoided and transmitted light maximized. Even if the coupling efficiency was lower than the optimal SPR condition, the increased light reaching the other side near the Brewster condition would help offset this. This may in practice correspond in the simplest case to a scattering correlation length similar to the pitch of a grating coupling mechanism but optimised to where most $p$ polarised light propagates through the film. In this situation, for transmitted light the loss is due to scattering, changing the angular resonance condition for a uniform film. This resonance should be strongest at or close to the Brewster angle. To explore this idea, we investigated whether normal sputtered films, which are rarely perfectly uniform, might demonstrate an additional weaker SPR at much smaller angles and appropriate thicknesses, closer to the orthogonal axis to the surface. If so this would relax the need for complex prism arrangements used to exploit SPR phenomena.

Two types of sputtered films were investigated: (1) commercially supplied Au films (PHASIS, Sweiss) firmly bound to borosilicate slides using a ~ 5 nm titanium layer (Au thicknesses, $t = 50$ nm) and (2) in-house fabricated Au films directly sputtered onto low grade borosilicate slides with varying thicknesses (Au $t = 30$, 45, and 60nm). A customised Kretschmann setup for measuring the SPR at the wavelengths of a HeNe laser (632.5 nm) was constructed where the normal SPR condition was monitored using a prism (resolution was ~ 1°). To confirm the near orthogonal excitation, the prism can also be removed reducing the amount of scattering from various prism interfaces.

In practice we found only one commercial film out of the four tested (all 50 nm) which showed a near orthogonal launch whilst most of the in-house material above a thickness of 15 nm showed signs of excitation, although with significant variation between them making quantification difficult. Figure 1 shows the results for two similar films, one commercial (50 nm) and one in-house sputtered (45 nm), that demonstrated both the expected SPR ~ 43-45° and a near orthogonal (low angle), weaker SPR-like resonance around 5-20°. These are consistent with calculations for the SPR resonance using equation (2) and for Brewster angles derived from the real part of equation (1) only. Both samples show a much weaker and broadened feature at low angles corresponding to another SPR than that observed at the main excitation condition. Further, the 45 nm thick film (Figure 1 a & b) showed signs of two peaks consistent with two SPRs, one at each of the Au-dielectric interfaces (air and borosilica). The air side is expected to have resonances at smaller angles since the effective index is lower.

The ability to couple into both sides, which cannot normally be done at the main SPR condition since it relies on phase matching through an evanescent field, is highly suggestive of a scattering mechanism and is not dissimilar to a linear two dimensional array of spherical particles forming a film. Figure 2 shows atomic force microscope height images of each of the two films – in the in-house sputtered case, the film is clearly non-uniform and surface features are visible as well as near-periodic surface undulation. The average surface roughness is $Ra \sim 2.3$ nm. For the commercial film $Ra$ is specified at $Ra \sim 2$ nm whereas the AFM measurements show an average $Ra \sim 1.9$ nm. What is interesting is the observation of nanoscale cracking in the commercial film (seen as dark lines) which can be attributed to arising from some lattice and thermal expansion coefficient mismatch between the gold and titanium buffer layer used to improve attachment. The in-house film did not have such strong attachment and therefore no significant cracking. Whilst these samples were considered clean, it is possible that dirt and other material can be used to induce similar scattering. Whilst the differences in scattering components make meaningful quantification difficult these measurements indicate that surface roughness is present and therefore scattering will take place.

To confirm that an SPR is indeed present on the Au-air side when excited at a lower angle, spectra were taken on a Shimadzu UV-VIS-NIR spectrophotometer. In reflection mode the angle of incidence in the sample holder is $\sim 5°$, on the edge of the broad angular spectra observed. Together with some beam divergence, there should be sufficient light coupling into the film by the proposed mechanism. A polariser is used to select out $p$-polarised light to enhance the signal-to-noise. Varying results, from weak to strong, were obtained for all the films with the measured near orthogonal SPR. The commercial film was weakest in practice with only a slight signature of a resonance. Figure 3 shows the best results obtained which were for a 30 nm in-house sputtered film. Both water and toluene were used to demonstrate shifts in the SPR peak and from the plot of shift versus refractive index, RI = $n$, a wavelength sensitivity of $\Delta\lambda \sim 63$ nm/RIU is obtained. This is similar to the observed shift for the main SPR peak but somewhat less than the theoretically calculated value of 970 nm/RIU by Homola *et al.* [13] for a prism coupled system, suggesting scope for further improvements. In contrast to other methods required to excite the classical SPR, no prism is used allowing the excitation and spectral measurement to be conducted easily within a commercial spectrophotometer.

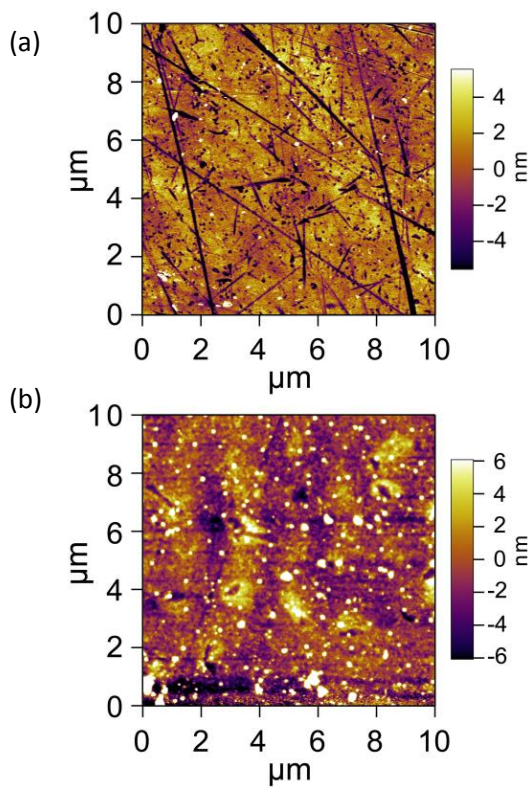

Fig. 2. Atomic force microscope (AFM) height images of (a) 50 nm thick commercially sputtered Au where it is difficult to observe any aggregated regions. Instead, nanoscale cracking is observed; and (b) 45 nm thick Au where the film is uneven and lumpy in appearance, with intermittent large scatterers. Also observable is an undulation with a quasi-period of $\sim 2\mu m$.

The underpinning coupling mechanism is thought to be scattering of near-Brewster transmitted light at interfacial roughness which exists in many sputtered films; the film surface roughness in Figure 2(a) can be treated like a series of planar aligned clusters of aggregated Au or individual particles. More work is necessary to verify and quantify the explanation as well as determine optimal conditions for scattering. For example, a 2-D array of gold clusters or particles approximating a film with periodic roughness may introduce grating-like coupling conditions to further enhance the process. The growth of plasmon resonances in self-assembled films of gold nanoparticles ($\phi \sim 6nm$) during layer by layer formation has been observed, implying that the nanoparticles retain their full angular character [14]. Given the visible dimensions of the lumpy regions shown in Figure 2(a), it is possible there is a scattering correlation length that essentially serves a similar role to a grating coupler with an effective pitch or alternatively retains broad angular incidence similar to individual particles. Either way, the conditions for classical SPR coupling are based around a uniform film with no scattering coefficients effectively ruling out the contribution of scattering centres that have complex local dipole orientations that do not resemble a uniform film.

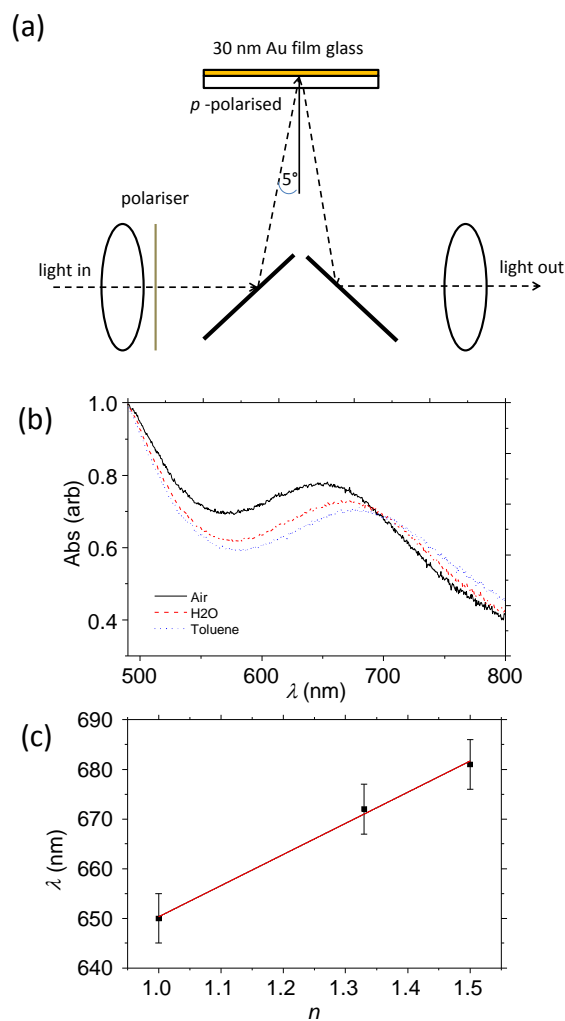

Fig. 3. Spectrophotometer measurements in reflection: (a) schematic of reflection configuration showing near orthogonal launch onto Au film where no prism is necessary; (b) the measured spectra for near-orthogonal excited SPR in the presence of air, water and toluene; (b) the peak wavelength shift from (a) as a function of refractive index, $n$.

The existence of such a condition within a standard routine deposition method greatly relaxes fabrication tolerances. If the Brewster condition is necessary, then the thickness criteria will be influenced in part by sufficient transmission of light suggesting some power dependence – this could account for some of the complex variation between samples we have observed. With the exception of one sample, the commercial films showed no detectable coupling, consistent with superior film uniformity. Given the interfacial confinement of surface waves, surface roughness must complicate the coupling mechanism and indeed the nature of the SPR – we suggest with sufficient surface roughness, localization of the SPR itself may be occurring within the sputtered films an example of Anderson localization [15] of surface waves. This is analogous to theoretical simulations where a metal-dielectric material, treated in terms of fluctuations in structural order, can generate localized surface plasmons [16] or, more closely, treated in terms of distributed scatterers which theoretically support Anderson localisation [17].

In summary, this work has demonstrated near-orthogonal coupling of light into Au films avoiding the need for traditional coupling methods. Whilst a scattering based mechanism points towards localized SPR generation through a scattering correlation length, and Anderson localization, more work is needed to unravel all the details. Nonetheless, the results have potential benefit for many device applications, greatly simplifying design criteria or offering alternative routes to exploiting SPR interrogation. Further, new routes to optimising and designing films for specific applications are possible. The work here is confined to sputtered Au but clearly other techniques to pattern gold, or any other material that supports surface waves, exist including depositing the material on pre-patterned surfaces, such as self-assembled fractal textures introduced by drying solvents [18] and self-assembly methods [14,19].

*Acknowledgements*: A. Karim and N. Tzoumis acknowledge *i*PL summer scholarships, Y. Tan acknowledges Prof. Baiou Guan and funding from Jinan University for her visit to *i*PL, and R. Patyk acknowledges support from CAPES in Brazil. P. Trimby from the Australian Centre for Microscopy and Microanalysis at The University of Sydney is acknowledged for assistance with Au deposition. The work is supported by ARC grants FT110100116 and FT110100225.